\documentstyle[epsfig,pre,preprint,aps]{revtex}
\tightenlines

\newcommand{\be}{\begin{equation}}

\newcommand{\ee}{\end{equation}}
\newcommand{\bs}{\begin{mathletters}} 
\newcommand{\es}{\end{mathletters}} 

\newcommand{\baa}{\begin{eqnarray}}
\newcommand{\eaa}{\end{eqnarray}}

\newcommand{\ba}{\bs\begin{eqnarray}}
\newcommand{\ea}{\end{eqnarray}\es}

\newcommand{\bt}[1]{\bs\label{#1}\begin{eqnarray}}
\newcommand{\et}{\end{eqnarray}\es}

\newlength{\www}

\newcommand{\bq}{\begin{equation}}
\newcommand{\eq}{\end{equation}}
\newcommand{\bqa}{\begin{eqnarray}}
\newcommand{\eqa}{\end{eqnarray}}
\newcommand{\ben}{\begin{enumerate}}
\newcommand{\een}{\end{enumerate}}
\newcommand{\bc}{\begin{center}}
\newcommand{\ec}{\end{center}}

\global\nulldelimiterspace = 0pt

\begin{document}
\draft

\title{Susy Virtual effects at LEP2 Boundary}

\author{
M. Beccaria$^{(1)}$, 
P. Ciafaloni$^{(1)}$, 
D. Comelli$^{(2)}$, \\
F. Renard$^{(3)}$,
C. Verzegnassi$^{(1)}$}

\address{
\vskip 0.5cm
${}^1$ 
Dipartimento di Fisica dell'Universit\`a di Lecce, I-73100, Italy,\\
Istituto Nazionale di Fisica Nucleare, Sezione di Lecce;\\
${}^2$
Istituto Nazionale di Fisica Nucleare, Sezione di Ferrara; \\
${}^3$
Physique
Math\'{e}matique et Th\'{e}orique, UMR 5825\\
Universit\'{e} Montpellier
II,  F-34095 Montpellier Cedex 5.}

\maketitle

\begin{abstract}

\noindent We examine the possibility that SUSY particles are light, i.e. have
a mass just beyond the final kinematical reach of LEP2. In this case, even if
light particles are not directly detected, their virtual effects are enhanced
by a ``close to threshold'' resonance in the s-channel. We find that this
resonant effect is absent in the case of light sfermions, while it is
enhanced in the case of light gauginos, since neutralinos and
charginos add coherently in some regions of the allowed parameter space. We
discuss this ``virtual-alliance'' in detail and try to examine the
possibilities of its experimental verification.
\end{abstract}

\newpage

One of the most interesting sectors of the experimental program at LEP2
\cite{LEP2} is the search for supersymmetric particles. 
In the specific case
of the lightest Higgs boson, these efforts are particularly supported by the
existence, within a large class of supersymmetric models~\cite{HiggsBound}, of
an upper bound of approximately 130-150 GeV that is not much beyond the final
kinematical reach ($\sim$ 100-110 GeV) of the accelerator. This has motivated
the rigorous and detailed study of the production mechanism and of its visible
manifestations that has been fully illustrated in several dedicated
references~\cite{Yellow}. Since the nature of the light Higgs couplings
with the SM gauge bosons and light fermions makes the detection of
virtual effects at one loop rather remote, not much effort has been
concentrated on this alternative possibility.  

For what concerns the remaining supersymmetric particles, the situation appers
to us slightly different. In fact, no definite rigorous upper bound exists on
their masses; one can only expect from reasonable arguments based on
``naturalness'' requests~\cite{Naturality} that a limit of a few hundred GeV
should not be violated.
On the other hand, the possibility of small but visible virtual effects is
not, a priori, unconceivable. In particular, the existence of SUSY particles
with a mass just beyond the LEP2 reach could be observed as a consequence of a
resonant enhancement of self-energies, vertices or boxes
due to the production threshold of couples of these particles in the
$s$-channel.  Note that this remark is far from being obvious because, in
principle, the separate enhancements coming from the different diagrams could
well interfere destructively and lead to an unobservable effect.

The aim of this paper is precisely that of showing that a specially favourable
situation is provided by the hypothetical existence of a light chargino with a
mass ``close'' to 100 GeV (in our analysis we shall assume that the
kinematical reach of LEP2 is 200 GeV; this assumption can be easily modified
if this turned out to be a pessimistic -or optimistic- input). In such a case,
the overall virtual contributions of self-energies, vertices and box origin
from chargino pairs to several observables will not be negligible. On top of
this, for a large sector of the parameter space of the considered model, an
important extra help will come from the simultaneous resonance of virtual 
neutralinos, whose effect will add coherently with that of the charginos.
This kind of ``virtual alliance'' would lead to small, but observable effects,
that we shall discuss here in some detail. As we shall show in the second
part of the paper, the effects would be completetly different in the case of
virtual contributions due to light sfermions since, owing to the zero spin of
the involved particles, the resonant mechanism is practically absent.
Therefore, the light chargino-neutralino contribution appears to be a
reasonably well identifiable one, in this special and favorable case.
We shall devote the first part of this short paper to a detailed
numerical analysis of this effect.  

To begin our investigation, we shall
choose the relevant observables that might be used as indicators of (small)
virtual SUSY effects. By definition, these observables must be those that will
be measured at LEP2 with the best experimental accuracy, and for which an
extremely accurate theoretical prediction within the SM is obviously
available. In practice, these requests select three optimal candidates
i.e. the muon production cross section $\sigma_\mu$, the related forward
backward asymmetry $A_{FB, \mu}$ and the cross section for hadronic (u, d, s,
c, b) production $\sigma_5$. For these quantities we shall assume the expected
experimental precision quoted in the recent dedicated Workshop~\cite{LEP2},
which amounts roughly to less than a relative one percent, keeping in mind
that this value might be (hopefully) improved.

To proceed in a rigorous and self-contained way, we have decided to evaluate
both the SM prediction and the SUSY virtual effect using the same
computational program. With this purpose, we have first carried on the SM
analysis using the semianalytic program PALM, that was illustrated in a
previous paper to which we defer for all the technical
details~\cite{Bosonic}. In a second step, we have added to the theoretical
PALM SM prediction, computed at the one loop level, the extra virtual SUSY
effects. This has been done in a consistent way by adding to the special,
gauge invariant combinations of self-energies, vertices and boxes that were
grouped in the SM calculation the corresponding SUSY
contributions. Technically speaking, this corresponds to adding systematically
\underline{finite} SUSY quantities since all contributions in our approach are
\underline{subtracted} at the $Z$ peak, $q^2 = (\mbox{ c.m. energy})^2 =
M_Z^2$. We do not insist here on these details since they can be already found
in Ref.~\cite{Bosonic} for what concerns the SM calculation; the discussion of 
the
SUSY virtual effects at one loop, at general $q^2$ values (here we only
consider the LEP2 boundary situation, $\sqrt{q^2}=200$~GeV ) will be given in
a more exhaustive forthcoming paper~\cite{NextPaper}.

The theoretical model that we have considered is the MSSM~\cite{MSSM}, whose
detailed discussion we omit. Our starting assumption has been the existence of
a light \underline{chargino} with a mass 
``just'' beyond the LEP2 reach. Obviously, this
input can (and will) be easily modified but we shall use it in a first
qualitative investigation.
We have assumed the GUT relation between the SU(2) $\otimes$ U(1) gauginos
soft mass parameters $M_1=\frac{5}{3}\tan^2\theta_w M_2$ to be 
satisfied~\cite{MSSM}.  
In our simplified approach we neglect left-right mixing in the 
sfermion mass matrices and we
take all physical slepton masses to be degenerate at a common value
$m_{\tilde{l}}$ and all squarks masses degenerate at $m_{\tilde{q}}$, and we 
shall
return on this point in the final comments.
We also take initial and final state fermions to be massless; 
this is justified at the c.m. energies that we consider since we 
cannot have the top as final state.
Fixing the mass of the lightest chargino, $M_2$ varies
accordingly as a function of the supersymmetric Higgs mass parameter 
$\mu$.
Gluinos are assumed to be so heavy that they are decoupled. This is justified
by recent bounds from hadronic colliders \cite{Experiment}; moreover we are
interested here in new physics coming from the weak SU(2) sector of
MSSM. Contributions coming from gluinos will be considered in a subsequent
paper \cite{NextPaper}.

Our approach is based on a theoretical description of the invariant scattering
amplitude at one loop of the process $e^+e^-\to f\bar{f}$ that uses, as
experimental input parameters, quantities that are measured (apart from the
electric charge $\alpha(0)$) on top of the $Z$ resonance, as discussed in
previous papers~\cite{Renard}. In terms of the differential cross section for
the corresponding process, this leads to the following expression:
\begin{equation}
\label{ref1}
{d\sigma_{ef}\over dcos\theta}(q^2, \theta)=
{3\over8}(1+cos^2\theta)\sigma^{ef}_1+cos\theta \sigma^{ef}_2
\label{E23}
\end{equation}
with \widetext
\begin{eqnarray}
\label{ref2}
\sigma^{ef}_1=&&N_f(q^2)({4\pi q^2\over3})\{ {\alpha^2(0)Q^2_f\over q^4 }
[1+2\tilde{\Delta}_{\alpha,ef}(q^2, \theta)] \nonumber\\ &&+2\alpha(0)|Q_f|
[{q^2-M^2_Z\over q^2((q^2-M^2_Z)^2+M^2_Z\Gamma^2_Z)}][{3\Gamma_e\over
M_Z}]^{1/2}[{3\Gamma_f\over N_f(M_Z^2) M_Z}]^{1/2} {\tilde{v}_e
\tilde{v}_f\over (1+\tilde{v}^2_e)^{1/2}(1+\tilde{v}^2_f)^{1/2}}\nonumber\\ &&
\times[1 + \tilde{\Delta}_{\alpha, ef}(q^2, \theta)) - R_{ef}(q^2, \theta)
-4s_ec_e \{{1\over \tilde{v}_e}V_{ef}^{\gamma Z}(q^2, \theta)+{|Q_f|\over
\tilde{v}_f} V_{ef}^{Z\gamma}(q^2, \theta)\}]\nonumber\\ &&+{[{3\Gamma_e\over
M_Z}][{3\Gamma_f\over N_f(M_Z^2) M_Z}]\over(q^2-M^2_Z)^2+M^2_Z\Gamma^2_Z}
[1-2R_{ef}(q^2, \theta)
-8s_ec_e\{{\tilde{v}_e\over1+\tilde{v}^2_e}V_{ef}^{\gamma Z}(q^2,
\theta)+{\tilde{v}_f|Q_f|\over (1+\tilde{v}^2_f)} V_{ef}^{Z\gamma}(q^2,
\theta)\}]\}\nonumber\\ && \label{E24}
\end{eqnarray}

\begin{eqnarray}
\label{ref3}
\sigma^{ef}_2= && {3N_f(q^2)\over4}({4\pi q^2\over3}) \{2\alpha(0)|Q_f|
[{(q^2-M^2_Z)\over q^2((q^2-M^2_Z)^2+M^2_Z\Gamma^2_Z)}] [{3\Gamma_e\over
M_Z}]^{1/2}[{3\Gamma_f\over N_f(M_Z^2) M_Z}]^{1/2}\nonumber\\
&&\times{1\over(1+\tilde{v}^2_e)^{1/2}(1+\tilde{v}^2_f)^{1/2}}
[1+\tilde{\Delta}_{\alpha,ef}(q^2, \theta)-R_{ef}(q^2, \theta)]
+{[{3\Gamma_e\over M_Z}][{3\Gamma_f\over N_f(M_Z^2)
M_Z}]\over(q^2-M^2_Z)^2+M^2_Z\Gamma^2_Z}\nonumber\\ &&\times[{4\tilde{v}_e
\tilde{v}_f\over(1+\tilde{v}^2_e)(1+\tilde{v}^2_f)}] [1-2R_{ef}(q^2,
\theta)-4s_ec_e \{{1\over \tilde{v}_e}V_{ef}^{\gamma Z}(q^2,
\theta)+{|Q_f|\over \tilde{v}_f} V_{ef}^{Z\gamma}(q^2, \theta)\}]\}
\nonumber\\ &&
\end{eqnarray}
\noindent
where $N_f(q^2)$ is the conventional color factor which contains standard QCD
corrections at variable $q^2$, and where the theoretical input in
Eqs.~(\ref{ref2}-\ref{ref3}) contains the partial \underline{leptonic
and (light) hadronic}
$Z$ widths $\Gamma_l$,
$\Gamma_f$ and the \underline{related}
weak effective angles $s_l^2$, $s_f^2$ ($v_{l, f}\equiv 1-4
s^2_{l, f}$) measured on top of the $Z$ resonance~\cite{Renard}. The functions
that appear in the brackets are defined as follows:

\begin{equation}
\widetilde{\Delta}_{\alpha,
ef}(q^2,\theta)=\widetilde{F}_{\gamma\gamma,ef}(0,\theta)-
\widetilde{F}_{\gamma\gamma,ef}(q^2,\theta)
\label{ref4}
\end{equation}

\begin{equation}
R_{ef}(q^2,\theta)= I_{Z,ef}(q^2,\theta)-I_{Z,ef}(M^2_Z,\theta)
\label{ref5}
\end{equation}

\begin{equation}
V^{\gamma Z}_{ef}(q^2,\theta)= \frac{\widetilde{A}_{\gamma
Z,ef}(q^2,\theta)}{q^2}- \frac{\widetilde{A}_{\gamma
Z,ef}(M^2_Z,\theta)}{M_Z^2}
\label{ref6}
\end{equation}

\begin{equation}
V^{Z\gamma}_{ef}(q^2,\theta)=
\frac{\widetilde{A}_{Z\gamma,ef}(q^2,\theta)}{q^2}-
\frac{\widetilde{A}_{Z\gamma,ef}(M^2_Z,\theta)}{M_Z^2}
\label{ref7}
\end{equation}

where \widetext

\begin{equation}
I_{Z,ef}(q^2,\theta)= {q^2\over q^2-M^2_Z}[\widetilde{F}_{ZZ,ef}(q^2,\theta)
-\widetilde{F}_{ZZ,ef}(M^2_Z,\theta)]
\label{E5}
\end{equation}

\begin{equation}
\widetilde{A}_{ZZ,ef}(q^2,\theta)=
\widetilde{A}_{ZZ,ef}(0,\theta)+q^2\widetilde{F}_{ZZ,ef}(q^2,\theta)
\label{E9}
\end{equation}

\begin{equation}
\widetilde{A}_{ZZ,ef} (q^{2}, \theta) = A_{ZZ} (q^{2}) - (q^2-M^2_{Z})[
(\Gamma_{\mu,e}^{(Z)} , v_{\mu,e}^{(Z)}) + (\Gamma_{\mu,f}^{(Z)} ,
v_{\mu,f}^{(Z)})+ (q^2-M_Z^2) A^{(Box)}_ {ZZ, ef} (q^{2}, \theta)]
\label{E10}
\end{equation}

\begin{equation}
\tilde{F}_{\gamma\gamma,ef} (q^{2}, \theta) = F_{\gamma\gamma,ef} (q^{2}) -
(\Gamma_{\mu,e}^{(\gamma)} , v_{\mu,e}^{(\gamma)}) -
(\Gamma_{\mu,f}^{(\gamma)} , v_{\mu,f}^{(\gamma)})- q^2 A^{(Box)}_
{\gamma\gamma, ef} (q^{2}, \theta)
\label{E6}
\end{equation}

\begin{eqnarray}  
{\widetilde{A}_{\gamma Z,ef} (q^{2}, \theta)\over q^2} &=& {A_{\gamma Z}
(q^{2})\over q^2} - ({q^2-M^2_{Z}\over q^2}) (\Gamma_{\mu,f}^{(\gamma)} ,
v_{\mu,f}^{(Z)})\nonumber\\ &&-(\Gamma_{\mu,e}^{(Z)} ,
v_{\mu,e}^{(\gamma)})-(q^2-M^2_{Z}) A^{(Box)}_ {\gamma Z, ef} (q^{2}, \theta)
\label{E7}
\end{eqnarray}

\begin{eqnarray}  
 {\widetilde{A}_{Z\gamma,ef} (q^{2}, \theta)\over q^2} &=& {A_{\gamma Z}
(q^{2})\over q^2} - {q^2-M^2_{Z}\over q^2} (\Gamma_{\mu,e}^{(\gamma)} ,
v_{\mu,e}^{(Z)})\nonumber\\ &&-(\Gamma_{\mu,f}^{(Z)} ,
v_{\mu,f}^{(\gamma)})-(q^2-M^2_{Z}) A^{(Box)}_ {Z\gamma, ef} (q^{2}, \theta)
\label{E8}
\end{eqnarray}

The quantities $A_{ij}(q^2)= A_{ij}(0)+q^2 F_{ij}(q^2)$ ($i, j=\gamma, Z$) are
the conventional transverse $\gamma,Z$ self-energies.
$A^{(Box)}_{\gamma\gamma,\gamma Z, Z \gamma, ZZ, ef}(q^2,\theta)$ are the
projections on the photon and $Z$ Lorentz structures of the box contributions
to the scattering amplitude ${\cal A}_{ef}$ and the various brackets
($\Gamma_{\mu},v_{\mu}$) are the projections of the vertices on the different
Lorentz structures to which $\widetilde{A}_{\gamma\gamma}$,
$\widetilde{A}_{ZZ}$, $\widetilde{A}_{\gamma Z}$, $\widetilde{A}_{Z \gamma}$
belong.  In our notations $A^{(Box)}_{\gamma\gamma}$ is the component of the
scattering amplitude at one loop that appears in the form $v^{(\gamma)}_{\mu,
e} A^{(Box)}_{\gamma\gamma} v^{(\gamma), \mu}_{f}$ where
$v^{(\gamma),\mu}_{e,f} \equiv -|e_0|Q_{e,f} \gamma^{\mu}$ is what we call the
photon Lorentz structure and analogous definitions are obtained for
$A^{(Box)}_{ZZ}$, $A^{(Box)}_{\gamma Z}$, $A^{(Box)}_{Z \gamma}$ with the $Z$
Lorentz structure defined as $v^{(Z),\mu}_{e,f} \equiv
-{|e_0|\over2s_0c_0}\gamma^{\mu}(g^0_{V,e,f}-g^0_{A,e,f}\gamma^5)$.

More details can be found e.g. in~\cite{Bosonic}. Here we only stress the fact
that all the previous quantities $\tilde\Delta_\alpha$, $R$, $V_{\gamma Z}$,
$V_{Z\gamma}$ are separately gauge-invariant and therefore their evaluation in
the SM can be performed without intrinsic ambiguities, leading to the
numerical results fully discussed in~\cite{Bosonic}.

\begin{figure}[htb]\setlength{\unitlength}{1cm}
\center{\epsfig{ file=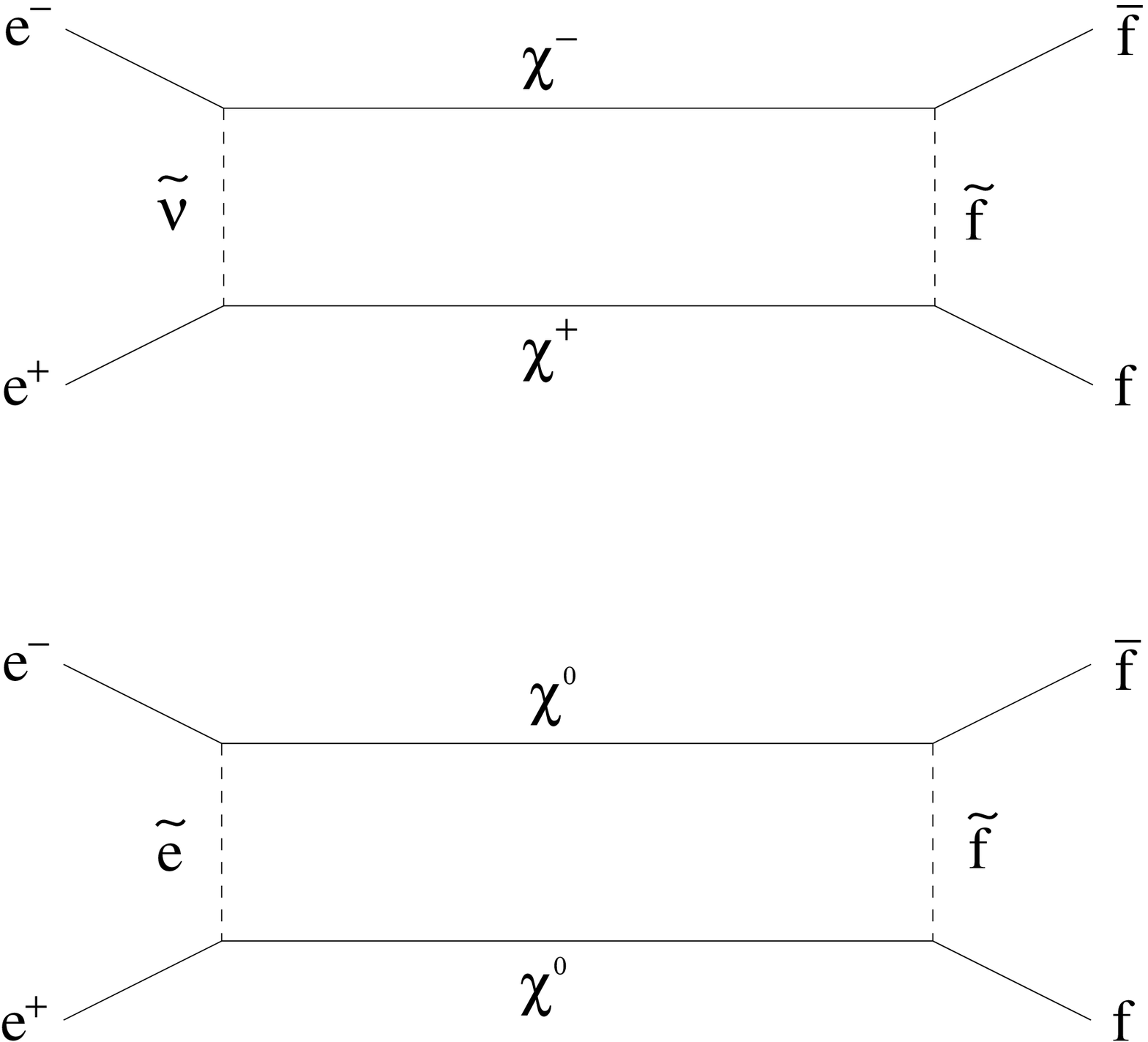 , width=11cm}}
\caption{Boxes}
\label{box}
\end{figure}

\begin{figure}[htb]\setlength{\unitlength}{1cm}
\center{\epsfig{ file=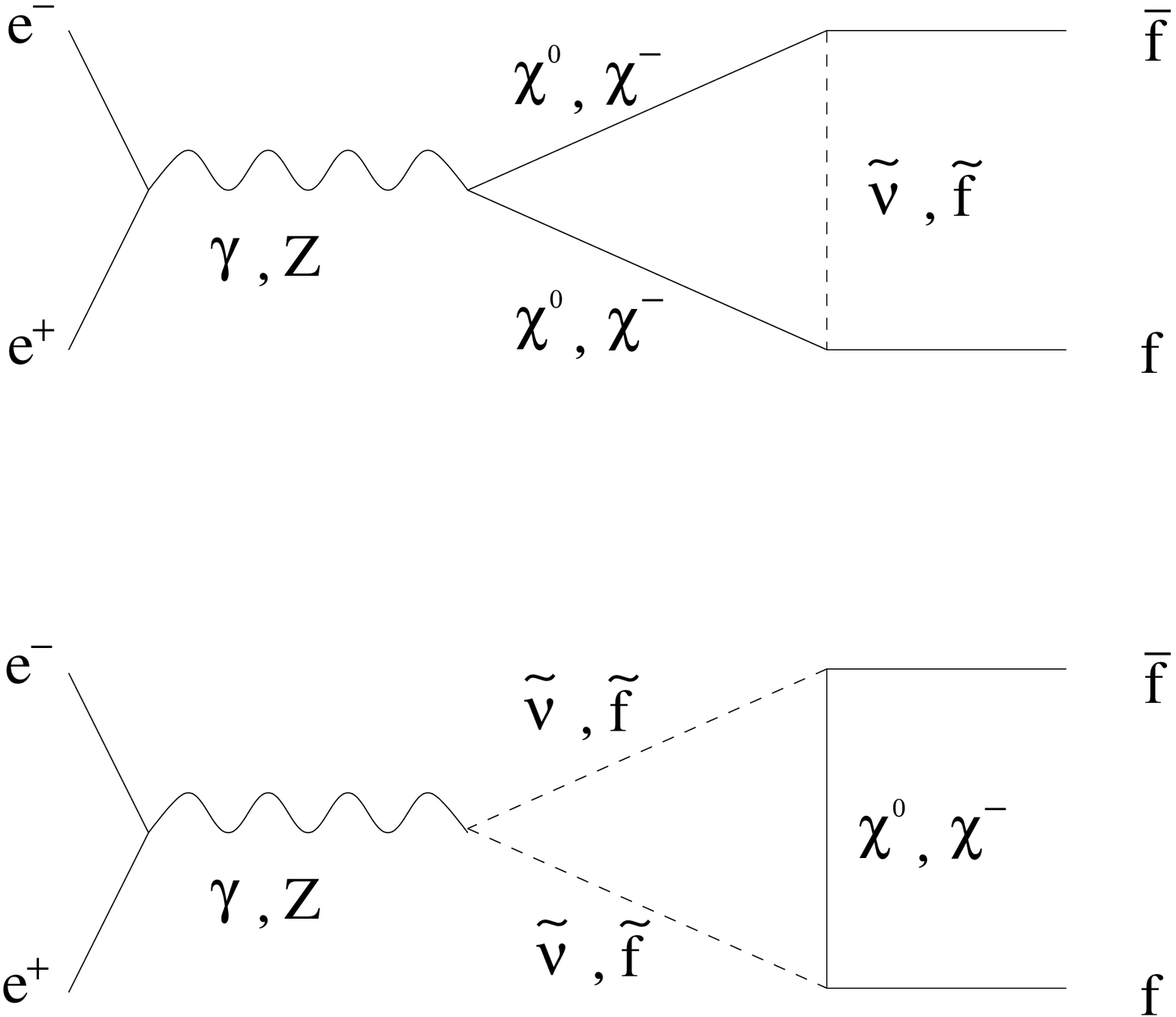, width=11cm}}
\caption{Vertices}
\label{vertex}
\end{figure}

\begin{figure}[htb]\setlength{\unitlength}{1cm}
\center{\epsfig{ file=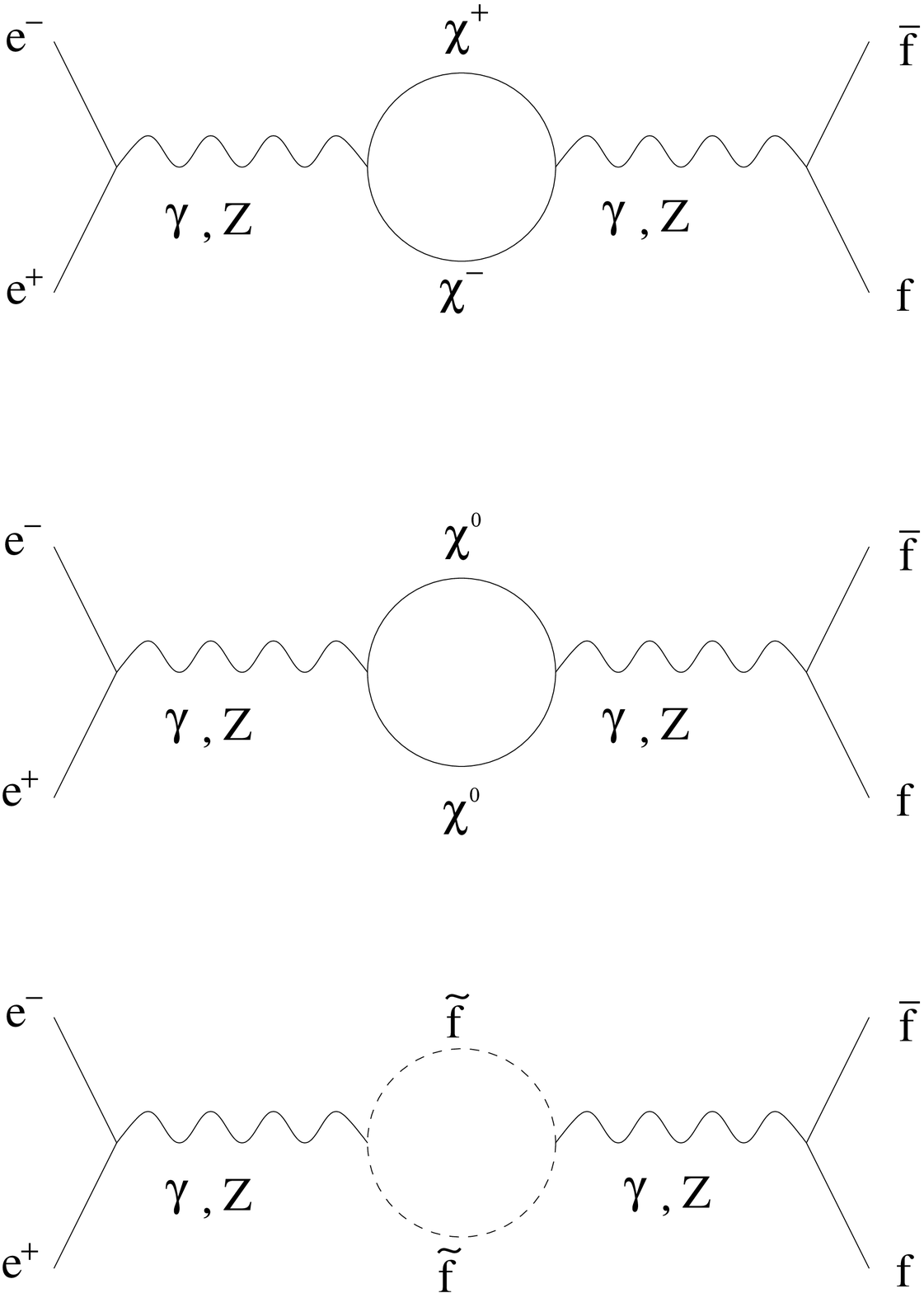, width=11cm}}
\caption{Self-energies}
\label{selfenergy}
\end{figure}

To compute the SUSY effect on the three chosen observables, we have calculated
the quantities $(\tilde\Delta_\alpha$, $R$, $V_{\gamma Z}$,
$V_{Z\gamma})^{SUSY}$. These are \underline{finite} contributions that are
generated by Feynman diagrams of self energy, vertex and box type. In
Figs.~(\ref{box},\ref{vertex},\ref{selfenergy}) we represent diagrammatically
some of the relevant graphs, omitting for simplicity other ones e.g. external
self-energy insertions.
As a technical comment, we would like to note that one could expect
various Lorentz-invariant Dirac structures to contribute to the amplitudes
under consideration, especially in the case of SUSY boxes that have a non
conventional structure with respect to the SM ones. 
However, due to a
``generalized Fierz identity'' that will be discussed in detail elsewhere
\cite{NextPaper}, it is possible to demonstrate that only four 
independent Dirac structures 
(i.e., $\gamma^\mu P_{L,R}\otimes \gamma_\mu P_{L,R}$ where $P_{L, R}$ are the 
chiral projectors) 
contribute in the massless external fermions case that we consider here.

Inserting the expressions of the SUSY contribution to
Eqs.~(\ref{ref4}-\ref{ref7}) into the general equations
(\ref{ref1})-(\ref{ref3}) and performing the angular integration by means of
the PALM program we have computed the overall (SM) and (SM+MSSM) values.
Although the program is able
to estimate ISR (Initial State Radiation) effects~\cite{Bosonic}, we have not
inserted for our present investigation at $\sqrt{q^2}=200$~GeV the discussion
of this kind of effects; we believe that for the purposes of this preliminary
investigation this attitude can be safely tolerated.

We have first considered a case in which the light chargino mass is fixed at
105~GeV, the sleptons physical masses are equal to 120~GeV and the squarks
physical ones are assumed to be 200~GeV.  We set $\tan\beta=1.6$, and verified
that varying it from 1.6 to 40 does not produce any appreciable change. With
this choice, we computed the \underline{relative} SUSY shifts on the three
chosen observables ${\cal O}_i$, $\Delta^{\rm SUSY}{\cal O}\equiv \frac{{\cal
O}^{\rm SUSY}-{\cal O}^{\rm SM}}{{\cal O}^{\rm SM}}$ (${\cal O}_{1, 2,
3}=\sigma_\mu, \sigma_5, A_{FB, \mu}$).

\begin{figure}[htb]\setlength{\unitlength}{1cm}
\center{\epsfig{ file=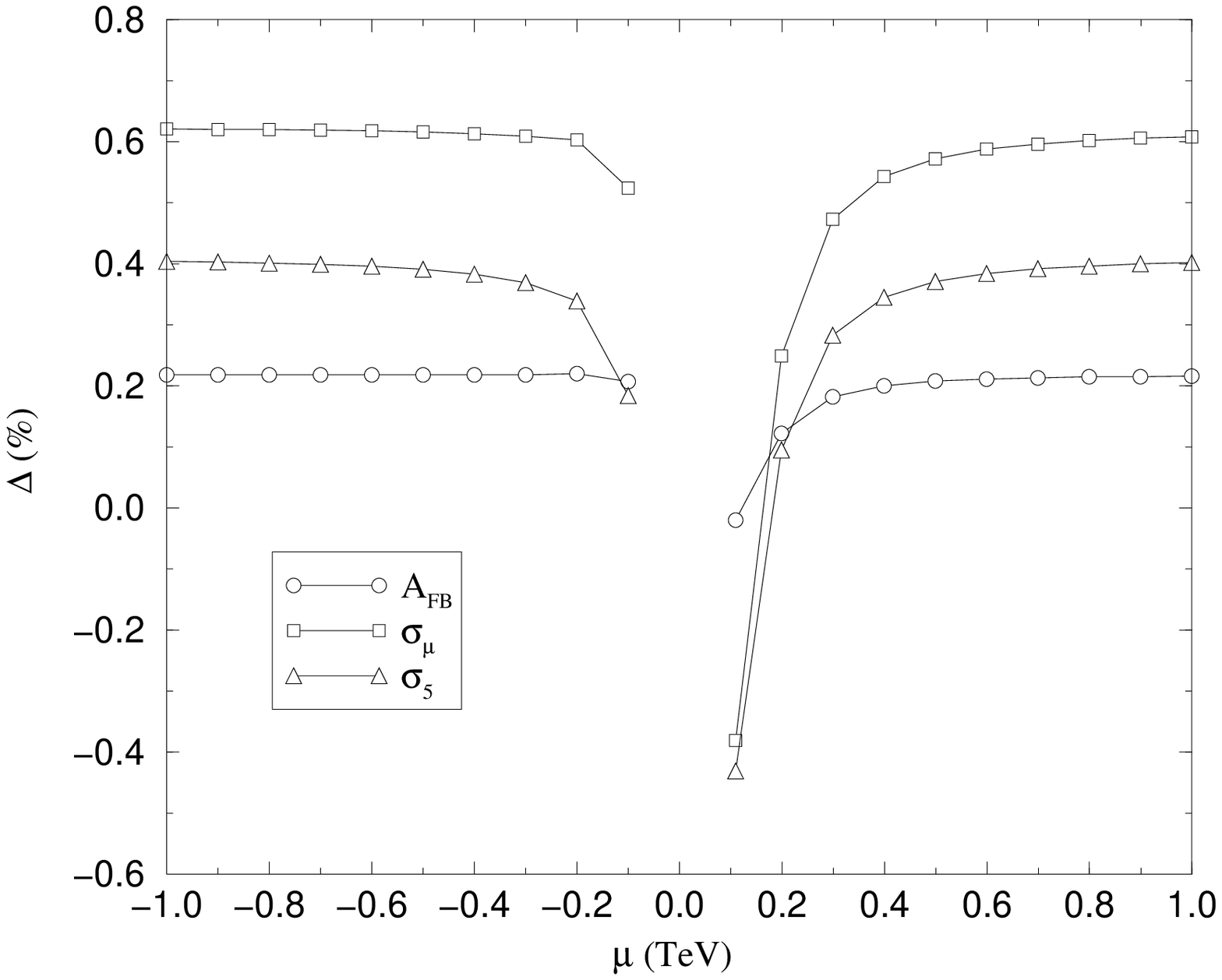,
width=12cm}}
\caption{SUSY effects on the three considered observables with the mass of the
lightest chargino fixed at 105 GeV and $\tan\beta=1.6$.  $m_{\tilde{q}}$ is
fixed at 200 GeV and $m_{\tilde{l}}$ at 120 GeV.}
\label{scan105}
\end{figure}

Fig.~(\ref{scan105}) shows the variations of the relative effects on the
observables when $\sqrt{q^2}=200$~GeV and 
$\mu$ varies in its allowed range. One sees that the size of
the SUSY contribution to the muon asymmetry remains systematically negligible, 
well below the six-seven permille limit that represents an
optimistic experimental reach in this case~\cite{LEP2}. The weakness
of this effect is due to two facts, the dominance of the photon
contribution in $\sigma^{e\mu}_1$ and of the photon-$Z$ interference
in $\sigma^{e\mu}_2$, and a subsequent accidental cancellation 
between $\tilde{\Delta}_{\alpha,e\mu}$ and $R_{e\mu}$ in the resulting
$\tilde{\Delta}_{\alpha,e\mu}$ + $R_{e\mu}$ contribution 
to $A_{FB, \mu}$.
On the contrary, in
the case of the muon and hadronic cross sections, the size of the effect
approaches, for large $|\mu|$ values, a limit of six permille in $\sigma_\mu$ 
and 
four permille in $\sigma_5$ that
represent a conceivable experimental reach, at the end of the overall LEP2 
running
period.

\begin{figure}[htb]\setlength{\unitlength}{1cm}
\center{\epsfig{ file=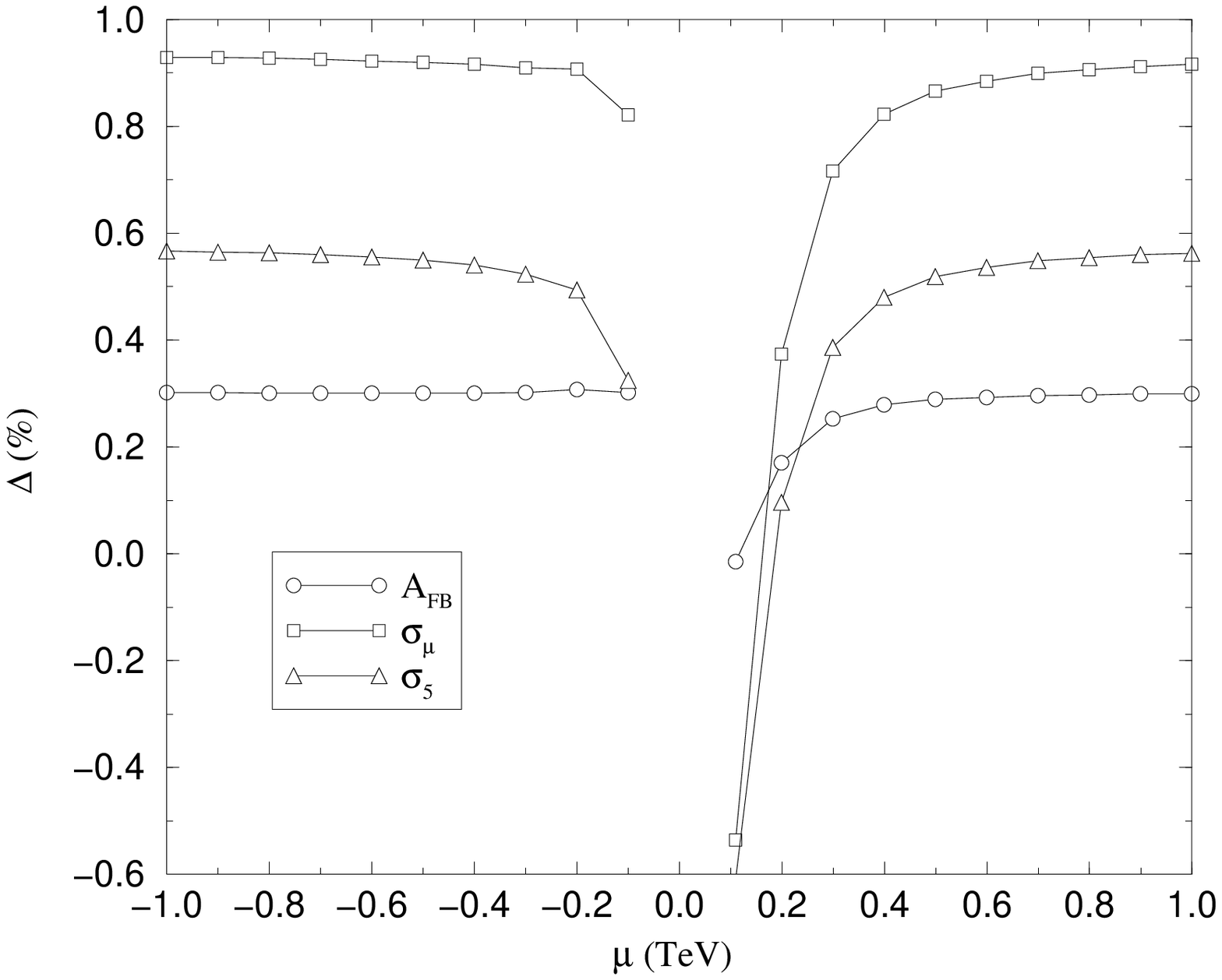,
width=12cm}}
\caption{SUSY effects on the three considered observables with the mass of the
lightest chargino fixed at 100 GeV and $\tan\beta=1.6$.  $m_{\tilde{q}}$ is
fixed at 200 GeV and $m_{\tilde{l}}$ at 120 GeV.}
\label{scan100}
\end{figure}

This explains in fact our choice of the value $M_{\chi^+_{light}}=
105$~GeV with LEP2 limit at 200~GeV; other couples of the light chargino mass
and of the LEP2 limit separated by a larger gap would produce a smaller
effect, i.e. an unobservable one. On the other hand, smaller gaps (e.g. a
lighter but still unproduced chargino \underline{or} a larger LEP2 limit)
would increase the effect, as one see in Fig.~(\ref{scan100}), towards the one 
percent 
values that appear to be experimentally realistic.

\begin{figure}[htb]\setlength{\unitlength}{1cm}
\center{\epsfig{ file=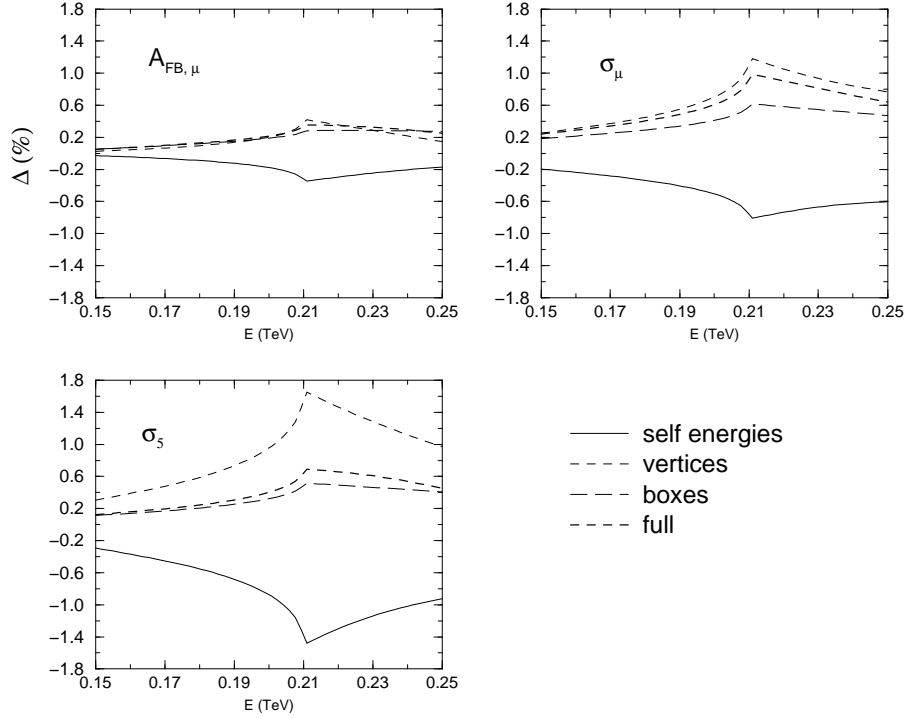, width=12cm}}
\caption{Heavy sfermions-light chargino scenario.  Selfenergy, box and vertex
SUSY effects on the three considered observables as a function of the
c.m. energy with a high $|\mu|$ value. The mass of the lightest chargino is
fixed at 105 GeV. The other parameters are: $m_{\tilde{q}}$=200 GeV,
$m_{\tilde{l}}$=120 GeV, $\tan\beta=1.6$.  }
\label{lightchar}
\end{figure}

\begin{figure}[htb]\setlength{\unitlength}{1cm}
\center{\epsfig{ file=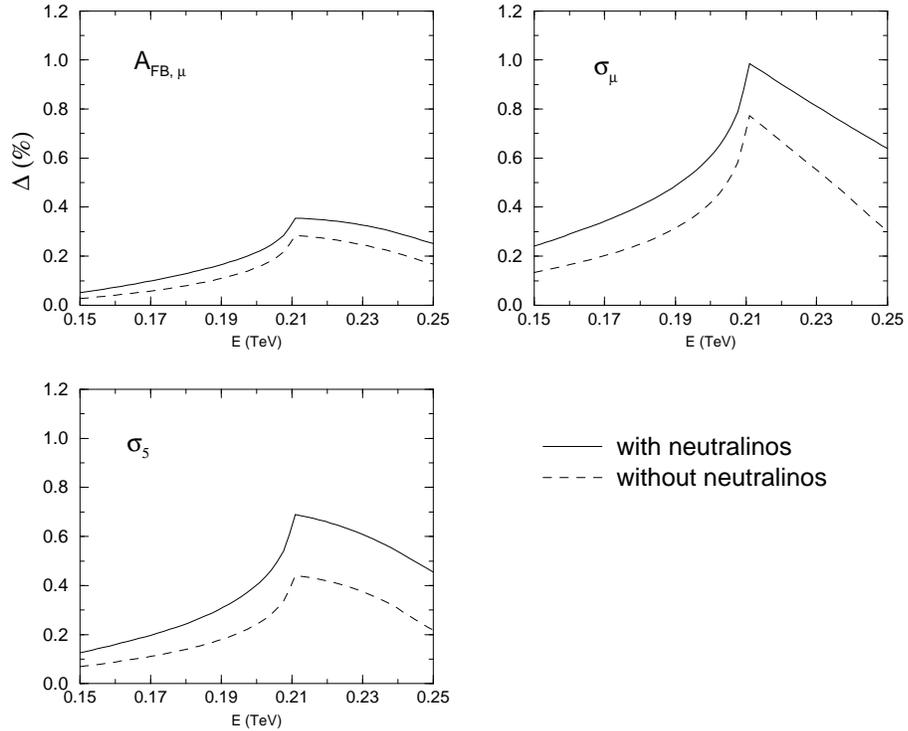,
width=12cm}}
\caption{Total SUSY effects on the three considered observables with and
without neutralinos contribution. The parameters are the same as in the
previous figure.}
\label{neutralino}
\end{figure}

Let us now discuss the qualitative features of the results that we obtain.  
As one
sees from Fig.~(\ref{lightchar}), the one loop SUSY effects have different
signatures. Those of ``oblique'' (universal) type, corresponding to
self-energies, have a negative effect on all three observables; those of
non-universal type (vertices and boxes) lead to a positive one in all the
three cases. Now, when $|\mu|>>M_2$ we have a light ``gaugino like'' chargino
of a (fixed) mass 105 GeV $\approx M_2$, 
and a heavy ``higgsino like'' chargino of 
a mass of the
order of $|\mu|$ itself. At the same time, 
in the neutralino sector the situation is very similar, 
with two
heavy ``higgsino like'' neutralinos, and two light  ``gaugino like''
neutralinos of masses $M_1$ and $M_2$.
Then, 1 chargino + 1 neutralino are
``gaugino like'' and have roughly the same mass of order 
$M_2\approx 105$ GeV ``resonating'' coherently
in the vertex, box and self-energies
contributions.  This situation, which we call of ``virtual alliance'', is made
evident in Fig.~(\ref{neutralino}) where neutralinos are seen to contribute
for about 25 \% ot the total signal.  
Note that an important contribution to the
overall effect in the chosen configuration is that coming from the SUSY
\underline{boxes}\cite{neut}.

The opposite happens when $M_2>>|\mu|$. In this situation we have light
``higgsino type'' charginos and neutralinos, with masses of the order
$|\mu|\approx$ 105 GeV, and heavy ``gaugino type'' charginos and neutralinos. 
Since higgsinos are decoupled from
massless fermions, their contribution to boxes and vertices disappear and the 
overall signal is consequently weakened (see Fig.(5)).

\begin{figure}[htb]\setlength{\unitlength}{1cm}
\center{\epsfig{
file=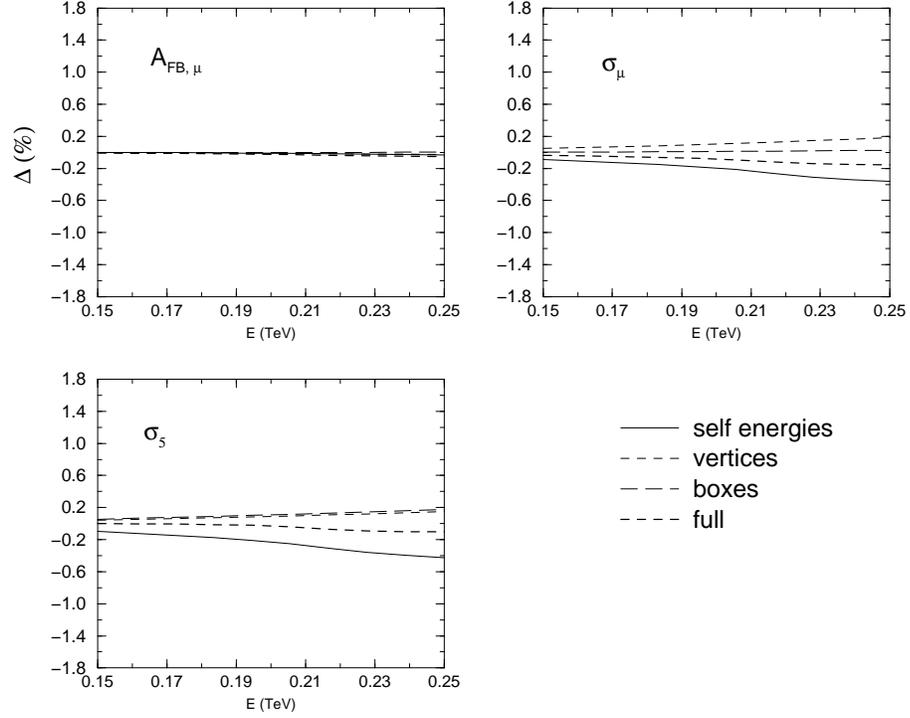, width=12cm}}
\caption{Light sfermions-heavy chargino scenario.  Selfenergy, box and vertex
SUSY effects on the three considered observables as a function of the
c.m. energy with a high $|\mu|$ value. The mass of the lightest chargino is
fixed at 300 GeV. The values of physical sfermion masses are:
$m_{\tilde{l}}$=105 GeV, $m_{\tilde{q}}$=200 GeV. }
\label{heavychar}
\end{figure}

Let us now consider a different situation, where the lightest chargino is
``heavy'' and decoupled, setting its mass equal to 300 GeV, and assuming that
all sleptons are now ``light'' (i.e.  $m_{\tilde{l}}=$ 105 GeV). The analogue
of Fig. (\ref{lightchar}) is then represented in Fig. (\ref{heavychar}).  As
one sees from the figure, the signal has now almost completely
disappeared. This fact can be qualitatively interpreted as a disappearance of
the ``quasi resonance'' chargino-neutralino contributions, not compensated by
analogous sleptons terms.  The reason is the fact that \underline{spinless}
particles, differently from spin 1/2 particles, have to be produced 
because of angular
momentum conservation, in a l=1 angular momentum state.  This causes a
relative ``threshold'' p-wave depression factor $\approx q^2-4
m_{\tilde{l}}^2$ in the spinless case, which washes out the threshold
enhancement.  Note also that, since we are not 
considering final electron-positron states, 
we don't have any box contributions with sfermions pairs in the s-channel
(see Fig. (\ref{box})).

Another important comment is related to our choice of using a ``Z-peak
subtracted'' representation. This has the consequence 
that all the \underline{energy independent} 
new physics contributions that can be reabsorbed in the Z-peak input quantities
($\Gamma_f, \sin^2\theta_{eff},...$) do not affect our final result.
Such is the case for all those 
values of sfermions splittings and/or mixings that contribute to
the $\Delta\rho$ parameter . These contributions 
are automatically reabsorbed when we replace $G_\mu$ by $\Gamma_l$ as 
theoretical input. They are, though, taken into account by the experimental 
error
on our theoretical input, in this case $\Gamma_l$. As exhaustively discussed 
in~\cite{Bosonic}
this would generate a strip of theoretical error in our prediction of the one 
permille size
well below the considered LEP2 experimental accuracy.

Note that, as a consequence of this ``LEP1 based'' approach, all our
residual subtracted theoretical  one-loop combinations of self energies,
vertices and boxes are finite and thus separately computable. In a
forthcoming paper~\cite{NextPaper} we shall discuss in more detail a
dedicated numerical code (SPALM), that is already available upon request.

We should mention at this point that in a recent paper \cite{Hollik} a
calculation of virtual SUSY effect has been performed, that covers an
energy range from $200~GeV$ to the $TeV$ range. The approach followed
by the authors of Ref.\cite{Hollik} is different from ours,
particularly since the theoretical input parameters are different and
do not contain our LEP1 input. This makes a detailed comparison more
subtle, in particular for what concerns "relative" shifts when the
input parameters are different. Since the "virtual alliance" case that
we considered here has not been treated in Ref.\cite{Hollik}, we
postpone a complete and clean comparison of the two approaches to the
forthcoming paper \cite{NextPaper}.

In conclusion, we have seen that in the large $|\mu|$ configuration, a
delicate interplay exists between virtual SUSY contributions from
self-energies, vertices and boxes that might lead, for a conveniently light
chargino, to a small but visible effect. Our prediction is that the
signature of the effect is a positive shift of 
the muonic and hadronic cross sections. Given
the relative smallness of the signal, an important help comes from the light
neutralino, in particular from its box contribution that adds coherently to
that of the chargino, giving a 25 \% enhancement of the signal. This can be
interpreted as a kind of ``virtual alliance'', as we anticipated in the
abstract.
The observation of the predicted
simultaneous small excess in the two cross sections,
typically at the 1 \% level at most, could well be within the reach 
of a series of
dedicated LEP2 experiments.


\end{document}